\documentclass[12pt]{iopart}
\usepackage[english]{babel}
\usepackage{pdfsync}
\usepackage{subfigure}
\usepackage{graphicx}

\usepackage{amssymb}
\usepackage{iopams}
\begin{document}

\title[Extreme wave events at the initial stage of the modulational instability...]{Extreme wave events at the initial stage of the modulational instability in a forced and damped wave medium}

\author{E V Belkin, A V Kirichok$^1$, V M Kuklin}

\address{Kharkov National University, Institute for High Technologies, 4 Svobody Sq., Kharkov 61077, Ukraine}
\ead{sandyrcs@gmail.com}

\begin{abstract}
The modulational instability of waves in a medium under the action of an external monochromatic force and dissipation is considered. The model which describes the nonlinear stage of the modulation instability was constructed with using methods of so-called $S$-theory. Following this theory, we take into account only interactions between modes with slowly varying relative phases. Within framework of this model, it is shown that the waves satisfying the criterion of abnormality appears as a rule at the initial stage of the instability. With further development of the process, the abnormal waves can be also observed but their amplitudes decrease gradually with decreasing of the average wave height.
\end{abstract}


\submitto{\PS}
\maketitle

\section{Introduction}

In recent years, the attention of researchers was attracted to the phenomenon of large-amplitude short-lived waves observed in various nonlinear dispersive wave media. These waves was named as freaks or rogue waves. Of particular interest are the experimental observations of extreme ocean waves. A complete review on the various phenomena yielding to rogue waves in ocean can be found in the book \cite{Pelinovsky2009}.  Later, it has been found experimentally that freak waves can be generated in optical systems \cite{Solli2007}, \cite{Yeom2007}  and in space plasma \cite{Burlaga2007}. A possibility of generation of ion-acoustic freak waves in plasma with negative ions and generation of Alfv\'{e}n freak waves in astrophysical plasma was considered in \cite{Ruderman2010}.

Discussion about the possible mechanisms for the formation of freak waves has persisted for all last years. Now, the modulation instability is considered as main pretender on this role \cite{Kharif2003}. The phenomenon of modulational instability of finite amplitude wave was first investigated in \cite{Lighthill1965}, \cite{Benjamin1967}, \cite{Zaharov1967}. In the conservative media, that is in the absence of external energy sources and mechanisms of energy absorption, the final stage in the development of modulational instability is the formation of a train of solitary waves or regimes with peaking (collapse) \cite{Zaharov1972}, \cite{Zaharov1972c}. That's why there are persistent attempts to describe the development of modulational instability using modified in various ways solutions relating to conservative systems (see, for example \cite{Zemlyanaya2009}). However, this approach may not necessarily be constructive, since it does not take into account the rather active and long transition process of the formation of stable wave structures. What is more, the stability of these structures can be achieved only in narrow ranges of parameters.

Recently, Segur et al \cite{Segur2005} and Wu et al \cite{Wu2006} shown that dissipation stabilizes the modulational instability. In the wavenumber space, the region of instability shrinks as time increases. This means that any initially unstable mode of perturbation does not grow for ever. Damping can stop the growth of the side-bands before nonlinear interactions become important. Hence, when the perturbations are small initially, they cannot grow large enough for nonlinear resonant interaction between the carrier and the side-bands to become important. The amplitude of the side-bands can grow for a while and then oscillate in time. Kharif et al \cite{Kharif2010} considered the modulational instability of Stokes wavetrains suffering both effects of external forcing and dissipation. They found that the modulational instability depends on both frequency of the carrier wave and strength of the external force.

This work shows that the presence of energy flow through such an open system results in complete or partial suppression or delay the process of forming a train of solitary waves and creates a new metastable state. This state, at least at the initial stage of the instability (i.e. during only a few times greater than the inverse linear increment of instability), characterized by the appearance of short-lived splashes of modulation of the fundamental wave with abnormally large amplitude. The average amplitude of the fundamental wave therewith remains large enough. In the case of a large number of spatial modes in the excited spectrum, their amplitudes remain rather small in comparison with the amplitude of the fundamental wave and their interaction among themselves without the support of the fundamental wave can be neglected. At the initial stage of the instability, the width of the excited spectrum is rather wide and the process of interaction between side-band modes is accompanied in this case by so-called stimulated (i.e., governed by the fundamental wave) interference \cite{Kuklin2006}. Further development of the instability at large times leads to a narrowing of the spectrum and the formation of stable soliton-like large-scale formations that resembles structures obtained in \cite{Zemlyanaya2009}.

The method for describing the modulational instability in an open system was originally formulated in \cite{Suhl1958}. It was shown there that the pumping of spin waves by the homogeneous precession of magnetization ($ k_0 \to 0 $), that can be described by the Lighthill's equation, excites the symmetric spectrum satisfying the conditions of space-time synchronism of the form $ 2 \omega _0 = \omega (k) + \omega (-k) $ and $ 2k_0 = 0 = k-k$. Growing side-band spectrum acts back on the pump that leads to "freezing" of its amplitude at the threshold level and restriction of the instability.

Further improvement of the theory \cite{Schloman1960} takes into account the weaker interaction between excited low-amplitude modes, and the main contribution to this interaction is provided by pairs of waves symmetric with respect to the pump $ \omega (k)+\omega (-k) = \omega (k ') + \omega (-k ') $, which ensure that mentioned above conditions of space-time synchronizm are satisfied for all interacting modes. Later, Zaharov and co-authors (see detailed review \cite{{Zaharov1975}} and book \cite{Lvov1994}) have formulated the theory of the nonlinear stage of modulational instability based on this approach, which subsequently became known as $S$-theory. Within framework of this theory they come to description in the language of correlation functions $\langle A_K A_{k'}^{*} \rangle =n_K \Delta (k-k')$ and $\langle A_K A_{k'} \rangle =\sigma _K  \Delta (k+k')$, taking into account the effect of total coupling (correlation) of phases  $\varphi _K $ � $\varphi _{-k} $ for modes synchronously interacting with the uniform pump field and representing $\sigma _K =n_K\exp \{ -i\psi _K \} $.

Among important results of this extremely constructive $S$-theory are the discovered dominant mechanism of instability saturation, which consist in back action of the excited spectrum on the pump wave under condition of low above-threshold level, and growth of influence of the phase mismatch on the saturation of the instability with increasing above-threshold level  \cite{Zaharov1969}.

The model, considered below, is a continuation of the study \cite{Kirichok2008}, where we have analyzed the evolution of individual modes in the spectrum of plasma oscillations in the process of modulational instability.

%Thus, in the presence of an external source and the finite dissipation level, significant splashes of the envelope and the abnormally large waves appears as a rule at the initial stage of the modulational instability. The amplitude of each spatial mode of the modulation spectrum occurs much smaller than the amplitude of the fundamental wave, that allows to restrict the description of the system by only certain types of interaction.

In present work we study more carefully not only the dynamics of individual modes but also the spatial-temporal evolution of wave packets (e.g. waves and their envelope). In addition, we analyze some specific futures of the instability, in particular, the symmetry breaking of the excited spectrum during the progress of the modulational instability  in a medium with strong dispersion. This model, as shown below,  allows to detect the formation of abnormally high peaks of field modulation and the appearance of extremely high waves.

 \section{Mathematical model}

Consider, as an example, the modulational instability of externally driven wave in a medium with sufficiently strong dispersion and weak dissipation, which can be observed in plasma wave-guides, as well as on the water surface and other physical situations.  Weak dissipation, as noted in \cite{Segur2005}, provides necessary stabilizing effect on the modulation of fundamental waves, as an external source provides the permanent energy transfer to the wave motion.

%Let suppose that slow complex amplitude of wave perturbation is described by the externally driven and damped nonlinear Schr\"{o}dinger equation:                                                  
%\begin{equation} \label{0} 
%\frac{\partial A}{\partial t} =-\delta A-i\frac{\partial ^{2} A}{\partial x^{2} } -iA|A|^{2} +G   ,             
%\end{equation} 
%which arises in a variety of fields including plasma and condensed matter physics, nonlinear optics and superconducting electronics.
%Here $\delta $ is the linear dissipation coefficient, G is an external source, supporting the monochromatic fundamental wave on a frequency $\omega _{0} $ and wave-number $k=k_{0} $. 

It is significant that here, such as in the above discussed cases, the amplitude of externally forcing wave retains large during the instability development and greatly exceeds the amplitudes of the excited spatial spectrum. Therefore, in conditions of finite level of dissipation the side-band modes excited by the modulational instability will interact with one other only if this interaction is supported by the fundamental wave .
In other words, the dominant regime of interaction, at least in the early stage of the nonlinear regime of the instability is the interaction of modes symmetrically located in $\vec k$-space of the spectrum relative to the central mode of large-amplitude fundamental wave.

We use below the following dispersion relation that characteristic, as an example, of gravity waves on deep water \cite{{Debnath1994}}:
\begin{equation} \label{1)} 
\omega =\sqrt{gk} \left\{1+\frac{A^{2} k^{2} }{2} +...\right\}, 
\end{equation} 
where $g$ is some dimensional coefficient (for the gravity waves on deep water it is the acceleration of gravity). 
%Experimental data for ocean waves \cite{Schwartz1982} give us the following characteristics: the maximal steepness of progressive wave of permanent form is $H/\lambda = 0.13 - 0.14$, where $H\propto 2A$ and $\lambda $ are the vertical distance between the wave crest and the deepest trough preceding or following the crest and wave-length correspondingly. It follows from here that $Ak<1$.
%Denote the average wave amplitude as $|A_0 |$, the average wave height as $\bar{H}=2|A_0 |$. The gravity waves with $H=(2\div 3)\cdot 2|A_0 |$ are considered as extremely high. It follows from here that $(4\div 6){|A_0 |k_0}/{2\pi }  \propto 0.13$ and  it is easy to see that the width of spatial spectrum of the instability in this case is not so small in relation to the wave-number $k_0$ of the fundamental wave as in the Lighthill's model \cite{Lighthill1965}. So it is difficult to use here an expansion in the spatial scales as in \cite{Kirichok2008}.  

The equations for the field amplitude which satisfies the dispersion equation (\ref{1)}) can be represented as follows:  
\begin{eqnarray}  \label{eqn_2}
\nonumber \fl {\frac{\partial A_K }{\partial t} =-\delta A_K -i\sqrt{g(k_0 +K)} A_K }{-i\sqrt{g(k_0 +K)} \frac{(k_0 +K)^{2} }{2} \{ |A|^{2} A\} _K =}\\ \nonumber
{=-\delta A_K -i\sqrt{g(k_0 +K)} A_K -}{i\sqrt{g(k_0 +K)} \frac{(k_0 +K)^{2} }{2} \times } \\ 
{\times  \left({A_K {\left[ 2|A_0 |^{2} +2\sum _{\kappa\ne K,0}|A_{\kappa} |^{2}  +|A_K |^{2}\right] }{+A^*_{-K} \left[ A_0 ^{2} +\sum _{{\kappa} \ne K,0}A_{{\kappa} }  A_{-{\kappa} } \right]}} \right)}, 
\end{eqnarray} 
where $\lambda =2\pi /k_0 $ is the wave-length of the fundamental wave, $\delta $ is the linear dissipation coefficient.

Excluding the fundamental frequency $\omega_0=\sqrt{gk_0} $: 
\begin{equation}
\fl {A_K \propto \exp \{ -i\omega _0 t+i(k_0 +K)x+i\varphi _K \} \to } {A_K \propto \exp \{ +i(k_0 +K)x+i\varphi _K \} }, 
\nonumber
\end{equation} 
we can rewrite Eq.(\ref{eqn_2}):

\begin{eqnarray} \label{eqn_3}
\nonumber \fl{\frac{\partial A_K }{\partial t} =-\delta A_K -i[\sqrt{g(k_0 +K)} -\sqrt{gk_0 } ]A_K -}{i\sqrt{g(k_0 +K)} \frac{(k_0 +K)^{2} }{2} \times }\\ \nonumber
{\times \left({ A_K \left[2|A_0 |^{2} +2\sum _{K'\ne K,0}|A_{K'} |^{2}  +|A_K |^{2} \right]+}{A^*_{-K} \left[ A_0 ^{2} +\sum _{{\kappa} \ne K,0}A_{{\kappa} }  A_{-{\kappa} } \right] }\right) }. 
\end{eqnarray} 

Introducing real amplitudes and phases $A_K =|u_K |\exp(i\varphi _K)$  we have obtained the system of equations which describes the modulational instability in a medium with strong dispersion: 

\begin{eqnarray}
\label{eqn_4}
\fl \frac{\partial u_K }{\partial \tau } =-\delta u_K +(1+K)^{2.5}  \left[ u_{-K} u_0 ^{2} \sin\Phi_K + u_{-K} \sum _{{\kappa} \ne K,0}u_{{\kappa} }  u_{-{\kappa} } \sin(\Phi _{{\kappa} } -\Phi _K )\right] , 
\end{eqnarray} 
where  $\Phi _K=2\varphi _0 -\varphi _K -\varphi _{-K}$ is the total phase (or the phase of the instability channel).
A distinction needs to be drawn between modes with wave numbers $K$ and ${\kappa} $ and phases $\Phi _K $ and $\Phi _{{\kappa} } $. 

\begin{eqnarray}
\label{eqn_5}  
\nonumber \fl {\frac{\partial \varphi _K }{\partial \tau } =-\frac{2}{\alpha }(\sqrt{(1+K)} -1) -(1+K)^{2.5} \Biggl[2u_0 ^{2} + u_K ^{2}+2\sum _{K'\ne K,0}u_{K'} ^{2}   +}\\
\left.{+\frac{u_{-K} }{u_K } u_0 ^{2} \cos\Phi _K +\frac{u_{-K} }{u_K } \sum _{{\kappa} \ne K,0}u_{{\kappa} }  u_{-{\kappa} } \cos(\Phi _{{\kappa} } -\Phi _K )}\right], 
\end{eqnarray} 
where the mode frequencies are $\omega (K)-\omega _0 =\omega [k_0 (1+K)]-\omega (k_0 )=\sqrt{gk_0 (1+K)} -\sqrt{gk_0 } $. We use the following notations $k_0 \xi =\zeta $,  ${\omega _0t }/{2}=\tau /\alpha $, $\alpha =k_0 ^{2} |A_0 |^{2} $, $\tau =t\sqrt{gk_0 } (k_0 |A_0 |)^2 /2 $,  $K=(k-k_0 )/k_0 $, $\omega _0 tK^{2}/{4}=(\tau /2\alpha ) K^{2}$, $A_K /A_0 (\tau =0)=a_K =u_K \exp \{ i\varphi _K \} $, and also
\begin{eqnarray}
\nonumber \Delta _K =2\left(\sqrt{(1+K)} -1]+\sqrt{(1-K)} -1\right)/\alpha, \\
\nonumber {\rm P} _K=2(1+K)^{2.5} +2(1-K)^{2.5} -2. 
\end{eqnarray}
Note that radicands may be not expanded in further calculations. 

Equations for amplitude and phase of the fundamental wave take the form:                     
\begin{equation} \label{eqn_6} 
\frac{\partial u_0 }{\partial \tau } =-\delta u_0 -u_0 \sum _{{K \ne 0}}u_{K} u_{-K } \sin\Phi_K +G,    
\end{equation} 
\begin{equation} \label{eqn_7} 
\frac{\partial \varphi _0 }{\partial \tau } =-u_0 ^{2} -2\sum _{K\ne 0}u_K^{2} -\sum _{K \ne 0}u_K u_{-K} \cos\Phi_K .       
\end{equation} 
Here $G$ is an external source, supporting the monochromatic fundamental wave. The summing over $K$ can be replaced by summing over $K_{m} =m \Delta K/k_0 $, where $\Delta K=2K_{\max}/N$ and $m=\pm (1, 2,..., N)$ with $K_{\max} /k_0 =\sqrt{2} k_0 |A_0 |=\sqrt{2 \alpha}$, $N$ is the number of modes. Note that we consider direct external forcing  rather than parametric as in \cite{Kharif2010}. 

Equations \eref{eqn_4}-\eref{eqn_7} form a closed system that should be solved numerically.

%It can be shown that for $K/k_0 <1$ the linear increment of the instability $\gamma =\rm{Im}\Delta \omega $ is determined from the relation:                        
%\begin{equation}\label{eqn_8} 
%{-i\Delta \omega +\delta =-i\{ \alpha ^{-1} K+[1+\frac{3}{2}K^{2} ]|u_0 |^{2} \} \pm }  \\ 
%{\pm [-\{ -\frac{1}{2\alpha } K^{2} \} ^{2} -2\{ -\frac{1}{2\alpha } K^{2} \} [1+\frac{3}{2} K^{2} ]|u_0 |^{2} }{-8K^{2} |u_0 |^{4} \} ^{1/2} }, 
%\end{equation}    
%where the last term under the square root can be neglected. Obviously here $\Delta /2\approx -K^{2} /2\alpha $.

The wave profile in $\zeta$-space looks as follows:
 \begin{eqnarray} \label{eqn_9}
\nonumber \fl E=\exp \{ +i\zeta +i\phi _0 \} \{ u_0 +
\sum_{i>0} \left[u_{i} \exp \left(-i\frac{2\tau }{\alpha } (\sqrt{1+K} -1-K)+iK\zeta +i(\phi _{i} -\phi _0 )\right)\right.  + \\  {\left. +u_{-i} \exp \left(-i\frac{2\tau }{\alpha } (\sqrt{1-K} -1+K)-iK\zeta +i(\phi _{-i} -\phi _0 )\right)\right]}. 
\end{eqnarray} 
Here, the fundamental wave is at rest and modulation of moves.

Since the dispersion is sufficiently strong in the vicinity of the fundamental wave frequency, one can readily see from above equations the symmetry breaking in excitation of the Stokes and anti-Stokes part of the spectrum, which can significantly change the character of the modulation.

\section{Numerical results}

In order to analyze the wave height distribution (e.g. the distribution of vertical distances between the wave crest and the deepest trough preceding or following the crest ), we take a third of highest waves. Then we find the average height of all waves $U_{CP} $, average height of a highest third $U_{SWH}$ and the maximum wave height $U_{MAX}$ in consideration domain ($\zeta \subset L=2\pi /(\Delta K/k_0 )=\pi N/K_{m} =\pi N/\sqrt{2\alpha}$, where $\Delta K=2K_{\max}/N$, $\zeta =k_0 x$). Calculations were performed for 600 modes in the spectrum. The ratio of dissipation level $\delta$ to the maximum growth rate was chosen as 0.1 (e.g. $\delta =0.1$). In order to provide the unit amplitude of the fundamental wave at the initial stage of the instability we have also chosen the value of the external drive force as $G=\delta=0.1$ 

\begin{figure}[t]\center
\includegraphics[scale=1]{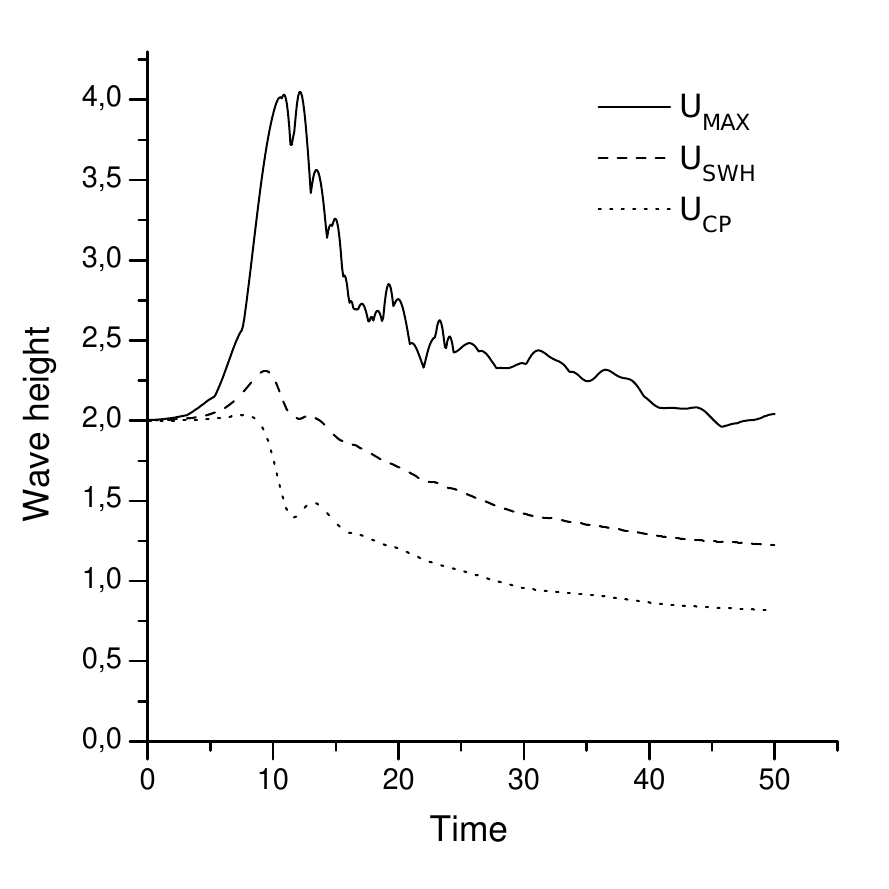}
\caption{Time evolution of wave height parameters: $U_{CP}$ -  the average height of all waves in the observation domain, $U_{SWH}$ - the average height of a highest third, $U_{MAX}$ - the maximum wave height \cite{Kuklin2009}.}
 \label{fig1}
\end{figure} 
 
Note that the criterion which defines the extremely high waves
\begin{equation} \label{eqn_10} 
U_{AG} >2U_{SWH}  
\end{equation} 
or something like this, should be used with caution because this criterion as usual is applied to statistics obtained on sufficiently large observation periods, but the highest amplitudes are observed at the initial stage of the instability.

\begin{figure}[t]\center
\subfigure[]{\includegraphics[width=0.47\textwidth]{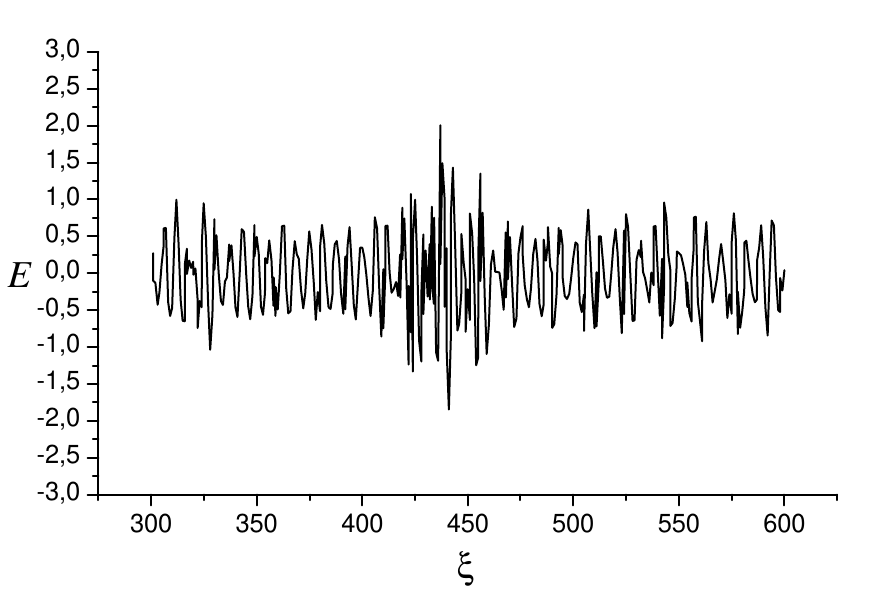}}  \hfill
\subfigure[]{\includegraphics[width=0.47\textwidth]{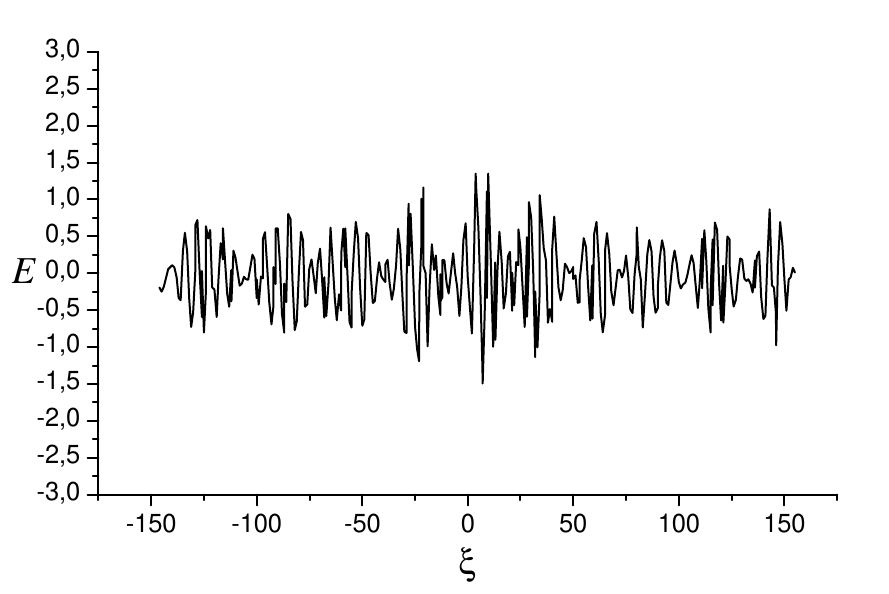}} \\

\caption{Wave profile (see \eref{eqn_9}) near extremely high waves for various realization of the process: (a) $\tau \propto 10$;  (b) $\tau \propto 30$}
 \label{fig2}
\end{figure}

Nevertheless, much smaller wave amplitudes or heights also satisfy this criterion in the regime of developed instability, since a decrease in both medium and large wave amplitudes is observed with time. Let give two examples of realizations where some waves satisfy the criterion \eref{eqn_10} (see Fig.\ref{fig2}).
                                                                                                                                                         
Analysis of observations and numerical simulation show \cite{Pelinovsky2009} that extreme wave events often occur in wave groups having the form of soliton-like structures.

\begin{figure}[t]\center
\includegraphics[scale=1]{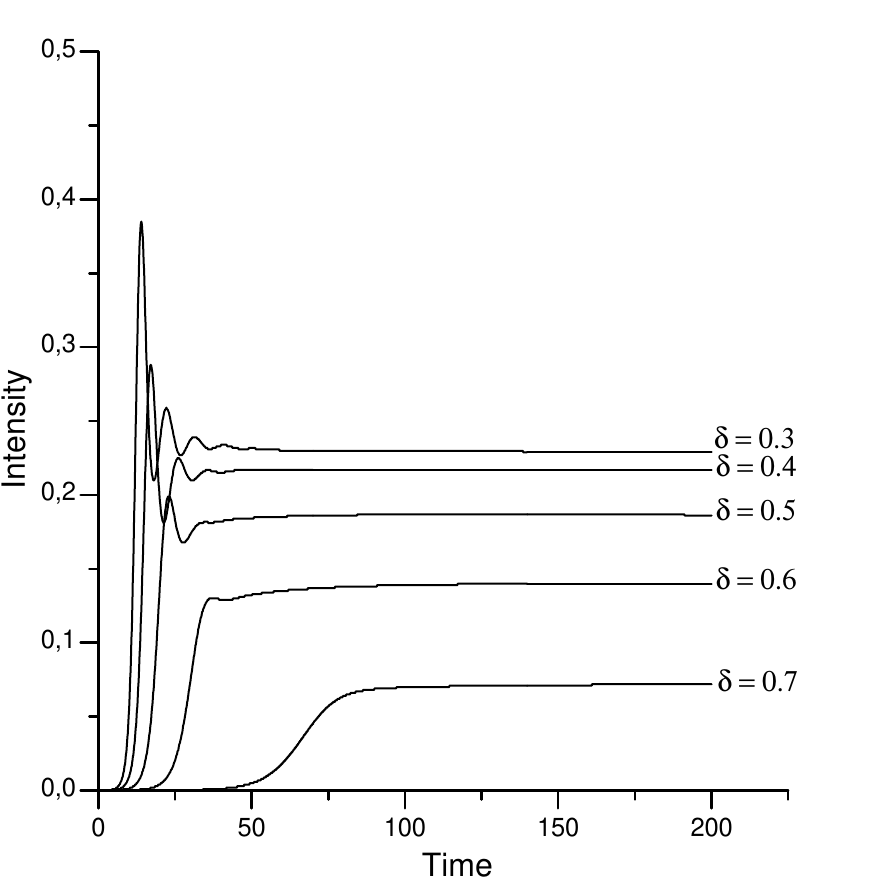}
\caption{The total intensity of side-band spectrum for different values of $\delta $ ($N= 100$)}
 \label{fig3}
\end{figure}

With decreasing dissipation level $\delta $ the processes of energy exchange between the side spectrum and the fundamental wave is amplified, which is clear seen in Figure \ref{fig3}.

It can be seen that the maximum of instability growth rate shifts to lower values of $K$ with decreasing amplitude of the fundamental wave at the initial stage of instability. Since the maximum growth rate corresponds to a value of the total phase $\Phi _K=2 \varphi _0 - \varphi _K - \varphi _ {-K} $ equal to $ \pi / 2 $, the majority of mode pairs with different values of $K$ are synchronized with the phase of the fundamental wave \cite{Kuklin1987} during instability progress. The fact that the total phase $ \Phi _K $ is focused near the $ \pi/2 $ sets the stage for almost one-type interaction of different mode pairs with the fundamental wave. This collective interaction of the growing spectrum modes with the fundamental wave and explains the nature of intense energy exchange between the fundamental wave and the spectrum at the initial stage of instability. Later, the spread of total phases grows and the energy exchange between the fundamental wave and the spectrum is attenuated. Owing to the fact that the phases of individual modes $ \varphi _K $ remain randomly distributed (in particular, no symmetry of the form $\varphi _K \ne \varphi _ {-K}$, $\varphi _K \ne - \varphi_ {-K } $), the instability spectrum synchronized with the fundamental wave forms a different interference structure for each realization. Nevertheless, the degree of coherence of excited modes achieves the maximum at the initial stage of the instability.
Incidentally, the phenomenon of synchronization of total phases in the spectrum of modulational instability in the presence of the symmetry $\varphi _K = \varphi _ {-K}$, usually leads to regimes with peaking \cite{Kuklin1988}. The absence of the phase symmetry in pairs of interacting waves could weaken the intensity of interference peaks and reduce the time of their existence.

\begin{figure}[t]\center
\includegraphics[scale=0.4]{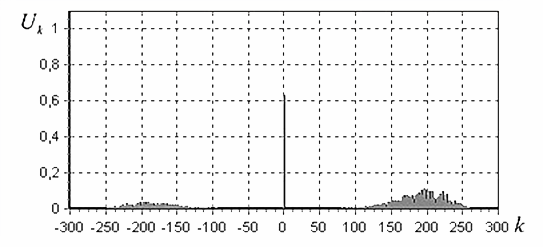}

(a)

\includegraphics[scale=0.4]{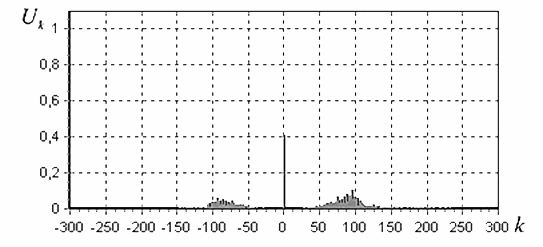}

(b)

\includegraphics[scale=0.4]{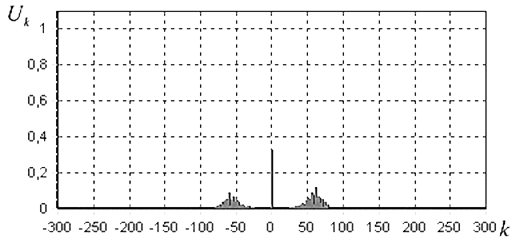}

(c)
\caption{The spectrum of instability at $\tau \propto 10$ (a), $\tau \propto 35$ (b), $\tau \propto 140$ (c)}
\label{fig4}
\end{figure}  

Considering the dynamics of instability spectrum, it is possible to detect the phenomenon of its noticeable shift in relation to the spectral domain of the linear instability. This shift can be explained by reduced amplitude of the fundamental wave. In addition, note that the amplitudes of individual modes of the spectrum  remain much smaller than the amplitude of the fundamental wave. It should be noted also the asymmetry of the instability spectrum relative to the fundamental wave due to strong dispersion and large enough modulational instability growth rate for fundamental waves.

\begin{figure}[t]\center
\includegraphics[scale=1]{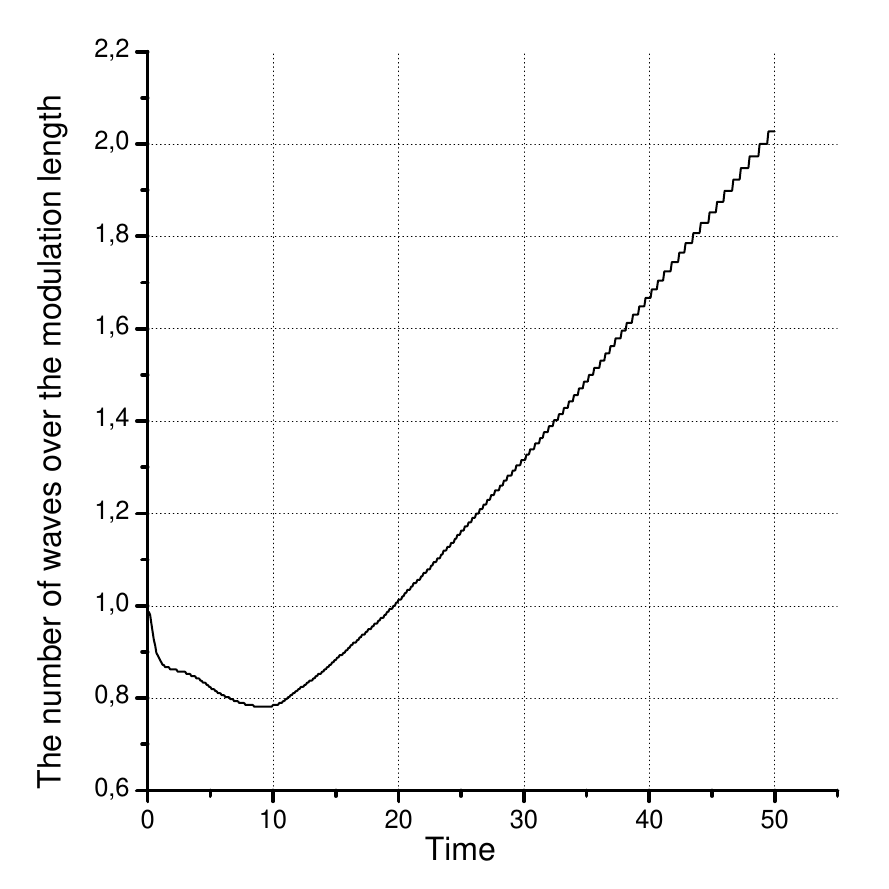}
\caption{Evolution of the relative characteristic modulation length with development of the instability.}
 \label{fig5}
\end{figure}

Analyzing the instability spectra one can see that the modulation length increases almost 3.5 times with development of the instability. Evolution of the relative characteristic modulation length is shown in Figure \ref{fig5}.

Dynamics of two-dimensional wave processes turns out to be similar. Thus, the number of waves on the modulation length at the initial stage of the instability is much less than at the later stages of the process.

Analyzing the probability of extreme wave events for different realizations of the process, we have found that one such wave appears among the 10000 waves, which is qualitatively consistent with the known observations of ocean waves.

\begin{figure}\center
\includegraphics[scale=0.5]{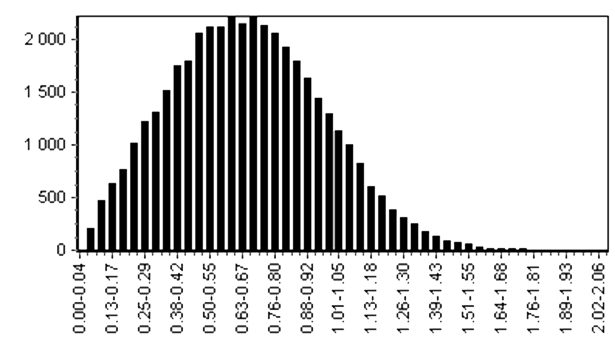}
\caption{Distribution of wave amplitudes in different realizations at the moment of $\tau \propto 10$.}
 \label{fig6}
\end{figure}

 \section{Conclusion}

The considered model shows that the appearance of waves with abnormally large amplitudes is characteristic of the initial stage of the modulational instability.  The average and maximum wave heights noticeably decrease with the development of the instability. However, even in the later case ($ \tau\propto 30$) abnormally high waves can be detected according to the criterion \eref {eqn_10}, although their amplitude is already one and a half - two times less than in the most interesting case ($\tau \propto 10$).

There are few waves fits at the modulation length in the initial stage of the process and one of which may be abnormally large in some realizations (our attention to this fact was called by E.Pelinovsky, \cite{Pelinovsky2009}). At the stage of developed instability, the number of waves for the modulation length increases to three or four times. The probability of abnormally high wave events in our numerical simulations is qualitatively consistent with the known observations of ocean waves. It should be noted that the total energy of wave motion is approximately conserved with reasonable accuracy in the domain under consideration (both in time and space) and the condition that the amplitudes of individual side-band modes remain much smaller of the fundamental wave amplitude is satisfied during the entire simulation time. The latter allows us to assume such a description of the modulational instability sufficiently correct.

The authors are grateful to Prof. E.A. Kuznetsov for his interest and helpful comments.

% \linenumbers



%% References


%\bibliographystyle{elsarticle-num}






% \end{thebibliography}

\end{document}